\documentclass[fleqn,10pt]{wlscirep}
\usepackage[utf8]{inputenc}
\usepackage[T1]{fontenc}
\usepackage{lineno}
\usepackage{xurl}
\usepackage{lscape}
\usepackage{rotating}
\newcommand{\rot}[1]{\begin{turn}{90}#1\enspace\end{turn}}

\usepackage{multirow}
\usepackage{placeins}
\usepackage[stable]{footmisc}
\usepackage{enumitem}
\usepackage{float}
\restylefloat{table}

\title{A maturity model for catalogues of semantic artefacts}

\author[1]{Oscar Corcho}
\author[2,3]{Fajar J. Ekaputra}
\author[4,5]{Ivan Heibi}
\author[6,7]{Clement Jonquet}
\author[8]{Andras Micsik}
\author[4,5]{Silvio Peroni}
\author[9,10]{Emanuele Storti}

\affil[1]{Ontology Engineering Group (OEG), Computer Science School, Universidad Politécnica de Madrid, Madrid, Spain}
\affil[2]{DPKM, Vienna University of Economic and Business (WU), Vienna, Austria}
\affil[3]{Data Science Research Unit, TU Wien, Vienna, Austria}
\affil[4]{Digital Humanities Advanced Research Centre (/DH.arc), Department of Classical Philology and Italian Studies, University of Bologna, Bologna, Italy}
\affil[5]{Research Centre for Open Scholarly Metadata, Department of Classical Philology and Italian Studies, University of Bologna, Bologna, Italy}
\affil[6]{MISTEA, University of Montpellier, INRAE \& Institut Agro, France}
\affil[7]{LIRMM, University of Montpellier \& CNRS, France}
\affil[8]{Department of Distributed Systems (DSD), Institute for Computer Science and Control (SZTAKI), Hungarian Research Network (HUN-REN), Budapest, Hungary}
\affil[9]{Department of Information Engineering, Polytechnic University of Marche, Ancona, Italy}
\affil[10]{European Council of Doctoral Candidates and Junior Researchers (Eurodoc), Brussels, Belgium}

\begin{abstract}
This work presents a \textit{maturity model} for assessing catalogues of semantic artefacts, one of the keystones that permit semantic interoperability of systems. We defined the dimensions and related features to include in the maturity model by analysing the current literature and existing catalogues of semantic artefacts provided by experts. In addition, we assessed 26 different catalogues to demonstrate the effectiveness of the maturity model, which includes 12 different dimensions (Metadata, Openness, Quality, Availability, Statistics, PID, Governance, Community, Sustainability, Technology, Transparency, and Assessment) and 43 related features (or sub-criteria) associated with these dimensions. Such a maturity model is one of the first attempts to provide recommendations for governance and processes for preserving and maintaining semantic artefacts and helps assess/address interoperability challenges.
\end{abstract}
\begin{document}

\flushbottom
\maketitle

\thispagestyle{empty}

\section*{Introduction}
With the advent of Open Data \cite{murray-rust_open_2008}, the Open Science movement \cite{unesco_unesco_2021}, and the FAIR Principles \cite{wilkinson_fair_2016} in the scholarly ecosystem, the role and need for storing, managing and sharing data grew significantly in academia. The General Data Protection Regulation (GDPR) has been an essential step in European data management in recent years. At least initially, the GDPR was the main responsible for scientists' fears of making their work impossible as data scientists. Indeed, one of the reasons for the introduction of the European Open Science Cloud (EOSC) has been to provide a safe European environment for data management compliant with the GDPR, to avoid the risk of European scientists starting to entrust all their data to foreign-owned/registered data servers to bypass European laws \cite{burgelman_politics_2021}.

In the EOSC, strategic relevance has been given (since the beginning) to address issues with implementing real interoperability among all the infrastructures, services and data researchers share in this ecosystem. Indeed, one of the most cited and used documents produced in projects implementing the EOSC is the EOSC Interoperability Framework, a report of the Interoperability Task Force of the EOSC Executive Board FAIR Working Group \cite{corcho_2021}. This report aimed to identify ``the general principles that should drive the creation of the EOSC Interoperability Framework (EOSC IF), and organises them into the four layers [...]: technical, semantic, organisational and legal interoperability'' \cite{corcho_2021}. In this framework, the objects that are referred to as the key components to enable the implementation of semantic interoperability are named \emph{semantic artefacts}. 

\subsection*{A definition for semantic artefact}
Previous studies used terms such as \textit{Knowledge Organization Systems} (KOS) \cite{zeng_2008} or knowledge artefact \cite{mcguinness_2003} to address semantic artefacts. A KOS has been adopted as a general term to encompass all schemes used to organise information and promote knowledge management, such as classification schemes, gazetteers, lexical databases, taxonomies, thesauri, and ontologies. These KOSs aim to underline the semantic structure of a domain, which needs to be embodied as web services to facilitate resource discovery and retrieval for either humans or machines. 

Other recent works defined the expression \textit{semantic artefact} as a machine-actionable and machine-readable formalisation of a conceptualisation enabling sharing and reuse by humans and machines \cite{corcho_2021, hugo_2020}. Semantic artefacts may have a broad range of formalisation, which include ontologies, terminologies, taxonomies, thesauri, vocabularies, metadata schemas, and standards \cite{corcho_2021,hugo_2020}. Semantic artefact was also strongly advised as an overarching term in the H2020 FAIRsFAIR project's “FAIR semantics” task. Despite the different forms of a semantic artefact, some works used blanket terms such as “ontologies” or “vocabularies and ontologies” \cite{jonquet_2018}. Moreover, semantic artefacts are serialised using various digital representation formats, e.g. RDF or OWL using XML, Turtle and JSON-LD \cite{hugo_2020}.

In the context of this article, we define a \emph{semantic artefact} as a machine-actionable formalisation (represented using appropriate formats and serialisations, including RDF and non-RDF standards) of a conceptualisation, enabling sharing and reuse by humans and machines. According to David \emph{et al.}\cite{david_converging_2023}, ``machine-actionability'' is defined as the property belonging to a type for which operations have been specified in a symbolic grammar. This entails that a machine-actionable definition is also machine-interpretable, i.e., can be related to semantic artefacts in a given context and, therefore, has a defined purpose, and machine-readable, i.e., is clearly defined by structural specifications.
Semantic artefacts may have a wide range of formalisations, from loose sets of terms, taxonomies, thesauri, (meta)data schemas, term mappings, and schema crosswalks to higher-order logic constructs, vocabularies, and ontologies.

\subsection*{Where semantic artefacts are preserved}
Often, these semantic artefacts are stored and shared using specific services called registries, libraries, repositories, catalogues, or terminology/vocabulary servers. Each provides a mixture of functionality –– ranging from simple metadata descriptions to advanced content-based services –– to facilitate finding, accessing, understanding and re-using of such semantic artefacts and enabling their long-term preservation.

The notion of \textit{ontology library} was introduced in \cite{ding_fensel_2001}, defined as “A library system that offers various functions for managing, adapting and standardizing groups of ontologies''. In addition, \cite{ding_fensel_2001} highlighted the importance of making such libraries easily accessible and offering efficient support for re-using existing relevant ontologies and standardizing them based on upper-level ontologies and ontology representation languages.

The terms “collection”, “listing”, or “registry” are also used to describe ontology libraries. All correspond to systems that help reuse or find ontologies by simply listing them (e.g., DAML or DERI listings) or offering structured metadata to describe them (e.g., FAIRSharing, BARTOC, Agrisemantics Map). Yet, those systems do not support additional services beyond the description of the items, e.g., a content analysis of the ontologies or a search index on the ontology content \cite{jonquet_2019}. A new concept introduced by Hartmann \emph{et al.}\cite{hartmann_et_al_2009} to cover these aspects is an \textit{ontology repository} with advanced features that enable searching, browsing, managing metadata, customizing, and mapping an application to query the contents of the ontologies. D'Aquin and Noy\cite{daquin_where_2012} and Naskar and Dutta\cite{naskar_ontology_2016} provide the latest reviews of ontology repositories.

By the end of the 2000s, the topic was of high interest as illustrated by the 2010 ORES workshop \cite{daquin_2010} or the 2008 Ontology Summit (\url{http://ontolog.cim3.net/wiki/OntologySummit2008}). The Open Ontology Repository Initiative \cite{baclawski_2009} aimed to create a joint infrastructure of ontology repositories through collaboration. At the time, the initiative utilized the NCBO BioPortal technology \cite{whetzel_2013}, which was the most advanced open-source technology for ontology management. Still, it was not yet available as a ``virtual appliance'' as it is today. Later, the initiative considered using the OntoHub \cite{codescu2017ontohub} technology for broader application, but it has since been discontinued.

Other recent works used to refer to a catalogue (that may or may not contain semantic artefacts) with terms such as ``repository'' \cite{corcho_2021,lin_2020} and ``registry''\cite{corcho_2021}, as well as hypernyms such as ``infrastructure'' and ``service''\cite{corcho_2021, ficarra_2020}. Additionally, the FAIRsFAIR project employed the term ``semantic registry'', which is defined as a ``catalogue that contains metadata about semantic artefacts''\cite{hugo_2020}.

While talking about catalogues, two other works provide relevant insights. Ficarra \emph{et al.}\cite{ficarra_2020} present the generic term \textit{open science infrastructures} and clarify that they are ``services, protocols, standards and software that the academic ecosystem needs to perform its functions during the research lifecycle''. Instead, Lin \emph{et al.}\cite{lin_2020} introduce the concept of \textit{trustworthy digital repositories} and provide an operational definition for them ``with a clear remit to actively preserve data in response to changes in both technology and stakeholder requirements''. Instead, the only works referring explicitly to semantic artefacts as the items contained in the catalogue are by Corcho \emph{et al.}\cite{corcho_2021} and, more recently, Jonquet \emph{et al.}\cite{jonquet_2023}.

Considering the status we have presented, there is a clear need to adopt an inclusive definition that, in principle, enables us to consider as a catalogue also web pages (e.g. \url{https://w3id.org/mobility}) with descriptive metadata of the semantic artefacts included in the catalogue in human-readable form. Thus, in the context of this article, we \emph{define a catalogue of semantic artefacts} as a dedicated web-based system that fosters the availability, discoverability and long-term preservation and maintenance of semantic artefacts.

\subsection*{Work setting and research question}
Two years ago, moved by the principles highlighted in the EOSC IF report, the EOSC Association (\url{https://www.eosc.eu}) promoted the creation of thirteen second-generation task forces (EOSC TFs, \url{https://eosc.eu/eosc-task-forces}), comprising selected experts from the members of the association, to create guidelines and tools to allow the development and deployment of the EOSC. These TFs have been grouped into four macro topics:

\begin{itemize}
    \item \emph{Metadata and data quality} - focussed on providing guidelines and tools for the discoverability, understanding, and reusability of research objects, and aiming at developing models and benchmarks to assess the data FAIRness and quality;
    \item \emph{Research careers and curricula} - dedicated to identifying the characteristics of the main stakeholders involved in the EOSC, i.e. researchers and data stewards, and possible rewarding systems to promote Open Science practices within the EOSC;
    \item \emph{Technical challenges} - directed at proposing approaches to foster authentication and authorisation procedures, augment the quality and FAIR-compliancy of research software, data and services, and their long-term preservation;
    \item \emph{Sustaining EOSC} - finalised at producing proposals for long-term financial sustainability of the EOSC and the monitoring of the TFs.
\end{itemize}

The first macro topic resonates with three distinct dimensions (each implemented through a different TF). Two of them have been dedicated to \emph{FAIR metrics and Data Quality} (\url{https://eosc.eu/advisory-groups/fair-metrics-and-data-quality}) and \emph{PID Policy and Implementation} (\url{https://eosc.eu/advisory-groups/pid-policy-implementation}) that have worked, respectively, on proposing metrics for measuring FAIRness and quality of data and on investigating approaches and development of persistent identifiers (PIDs) for emerging resource types to be used in scientific knowledge graphs according to the gaps listed in the EOSC Strategic Research and Innovation Agenda (\url{https://eosc.eu/sria-mar/}).

The third TF in the macro topic, which serves as the context of the present article, has been entirely dedicated to \emph{Semantic Interoperability} (\url{https://www.eosc.eu/advisory-groups/semantic-interoperability}) and aims at ensuring  ``that the precise format and meaning of exchanged data and information is preserved and understood throughout exchanges between parties'' \cite{corcho_2021}. The charter of this task force \cite{baumann_task_2021} and the subsequent working activities held in this context have concerned several aspects of semantic interoperability, including the identification of \emph{Semantic Interoperability Profiles} (SIPs, i.e. the list of resources used to address a specific semantic interoperability case study as defined by a community), mappings and crosswalks (for establishing relationships between elements in different models to achieve interoperability and data exchange), and the management and sharing of the primary tool for enabling semantic interoperability, i.e. semantic artefacts.

The work we describe in this article contributes to the ongoing effort by the EOSC Task Force on Semantic Interoperability to address interoperability challenges towards the vision of building the EOSC, as a Europe-wide shared data infrastructure based on the FAIR ecosystem of data and services. In particular, our work addresses the need to identify dimensions to assess the \emph{maturity} of catalogues of semantic artefacts. Catalogues are the critical system that enables and maximises the availability and discoverability of the semantic artefacts, thus acting as a keyholder and crucial component for implementing semantic interoperability.

Understanding the maturity of such catalogues is a crucial aspect to consider for envisioning how to enable and improve the long-term preservation of the semantic artefacts that permit semantic interoperability of systems. Indeed, a maturity model for assessing such catalogues would provide recommendations for governance and processes for preserving and maintaining semantic artefacts and help assess/address interoperability challenges. Additionally, improving these catalogues might help to enhance the data research life-cycle and other tools made available by research infrastructures. For example, the research infrastructure may warn the researcher about a new version of a semantic artefact, and help her find relevant mappings and adapt the research to the new version. Finally, the recent workshop organized by the Horizon Europe project FAIR-IMPACT on the governance of semantic artefacts \cite{ramezani_2023} has shown the importance of catalogues in the different governance models for semantic artefacts. 

This article presents the outcomes of an extensive analysis done by a group within the EOSC Task Force on Semantic Interoperability aiming at answering the following research question:

\begin{itemize}
    \item[] \emph{Which dimensions and features can be used to assess the maturity of catalogues of semantic artefacts?}
\end{itemize}

To answer this research question, we have gathered various definitions concerning catalogues storing and serving semantic artefacts (either at the metadata or data/content level or both). Then, by analysing the current literature on the topic, we have defined a model to measure, compare and evaluate available semantic artefact catalogues. We present this \textit{maturity model} as composed by several \textit{dimensions} in which catalogues could be compliant and/or enhanced. These dimensions facilitate the categorization of catalogues and the evaluation of their maturity. In addition, we analysed a collection of 26 semantic artefacts catalogues, aiming, on the one hand, at completing the maturity model by adding additional features (or sub-criteria) for each of the dimensions identified and, on the other hand, at showing how existing catalogues comply with such dimensions and sub-criteria. These features serve as a specification of different levels of compliance of a catalogue against the related dimension and provide a categorical view to assess the maturity of catalogues of semantic artefacts.

The target audience of our work and the maturity model presented here refers to at least the following users:

{\begin{itemize}
    \item Semantic artefact \emph{providers} may benefit from the model by understanding the criteria that make their artefacts more discoverable and usable. 
    \item \emph{Users} of semantic artefacts can leverage the maturity model to assess and compare different catalogues, gaining insights into the reliability and relevance of the offered artefacts. Both providers and users can rely on the model to select catalogues that align with their specific needs and expectations, e.g. level of accessibility, technical specifications or long-term usability of semantic artefacts.
    \item \emph{Developers} of semantic artefact catalogues can use the maturity model as a roadmap for designing and improving their applications. Understanding the criteria critical to users helps developers prioritise features and improvements that enhance the user experience and overall utility of the catalogue, ensuring they meet the criteria that users find valuable. This also enables developers to identify areas for continuous improvement in their catalogues through regular evaluation and benchmarking against industry standards and user expectations. This ensures they stay competitive and deliver a valuable resource for semantic artefact providers and users. Furthermore, this allows developers to communicate the strengths of their catalogues better to stakeholders, providers, and users, thus fostering trust and engagement.
\end{itemize}

The rest of the paper is structured as follows. Section `Results' presents our analysis, including the definition of relevant dimensions and the assessment of the selected catalogues. Section `Discussion' discusses the outcomes, outlines lessons learned, and sketches possible future developments. Finally, following the article guidelines of Scientific Data, Section `Methods' describes the selected material and our analysis methods, leading to the results previously described.

\section*{Results}\label{sec:results}

The section reports the dimensions and the related features we have identified by assessing 15 selected documents and 26 catalogues of semantic artefacts (Section `Dimensions and features'). We also characterise such selected catalogues of semantic artefacts according to the dimensions and their features (Section `Catalogues assessment'). 

\subsection*{Dimensions and features}
\label{sec:dimensions-features}

We have identified 12 dimensions that can be used to measure the maturity of the catalogues of semantic artefacts. These dimensions, summarised in Table \ref{t_mat_dimensions} with an indication of the documents that describe or refer to them, are listed in the following subsections. In addition, they are accompanied by several features (or sub-criteria) we have identified by analysing and harmonising the 26 catalogues of semantic artefacts selected.

\begin{table}[ht!]
\centering
\caption{Documents introducing the definitions of catalogues (that either may or may not refer to semantic artefacts explicitly) we have presented in Section `Introduction' and highlighting at least one maturity dimension identified by our analysis (Me: Metadata, Op: Openness, Qu: Quality, Av: Availability, St: Statistics, Pi: PID, Go: Governance, Co: Community, Su: Sustainability, Te: Technology, Tr: Transparency, As: Assessment).}
\begin{tabular}{||c || c | c c c c c c c c c c c c||} 
 \hline
\textbf{Document} & \textbf{Definition} & \textbf{Me} & \textbf{Op} & \textbf{Qu} & \textbf{Av} & \textbf{St} & \textbf{Pi} & \textbf{Go} & \textbf{Co} & \textbf{Su} & \textbf{Te} & \textbf{Tr} & \textbf{As} \\ [0.5ex] 
 \hline\hline
Alrashed \emph{et al.} (2021)\cite{Alrashed_2021} &  & x &  &  &  &  &  &  &  &  &  &  & \\
\hline
Benjelloun \emph{et al.} (2020)\cite{Benjelloun_2020} &  & x & x & x & x &  &  &  &  &  &  &  & \\
\hline
Bilder \emph{et al.} (2020)\cite{Bilder_2020} &  &  & x &  & x &  &  & x &  & x &  & x & \\
\hline
Brickley \emph{et al.} (2019)\cite{Brickley_2019} &  & x &  &  &  & x & x &  &  &  &  &  & \\
\hline
COAR and SPARC* (2019)\cite{Confederation_2019} &  &  & x &  & x &  &  & x &  &  & x & x & \\
\hline
Corcho \emph{et al.} (2021)\cite{corcho_2021} & x & x & x &  & x &  & x & x &  &  & x &  & \\
\hline
Cox \emph{et al.} (2021)\cite{Cox_2021} &  & x &  & x &  &  & x &  & x &  &  &  & \\
\hline
Ficarra \emph{et al.} (2020)\cite{ficarra_2020} & x &  & x &  &  &  &  & x & x & x &  &  & \\
\hline
French OS Committee (2019)\cite{French_2019} &  & x & x & x & x &  & x & x &  & x &  & x & x \\
\hline
Gregory \emph{et al.} (2020)\cite{Gregory_2020} &  &  &  & x &  &  &  &  & x &  & x &  & \\
\hline
Hugo \emph{et al.} (2020)\cite{hugo_2020} & x & x & x &  & x &  & x &  &  & x & x & x & \\ 
\hline
Jonquet \emph{et al.} (2019)\cite{jonquet_2019} & x & x & x &  & x & x & x &  & x & x & x & x & \\
\hline
Lin \emph{et al.} (2020)\cite{lin_2020} & x &  &  &  &  &  &  & x & x & x & x & x & \\
\hline
Skinner and Lippincott (2020)\cite{Skinner_2020a} &  &  & x &  & x &  &  & x & x & x & x & x & \\
\hline
Skinner and Lippincott (2020)\cite{Skinner_2020b} &  &  &  &  &  &  &  &  &  &  &  &  & x \\
\hline
\end{tabular}
\label{t_mat_dimensions}
\end{table}

\subsubsection*{Metadata (Me)}
The identification of the minimal set of metadata to describe the catalogue and its semantic artefacts. Huge importance is also given to using metadata standards and schemas (e.g., DCAT or Schema.org), adopting machine-readable formats, the documentation associated, and the licenses used to release the metadata. The six features identified in this dimension are:

\begin{enumerate}[label=\alph*)]
\item \textit{custom vocabulary} - custom metadata is used to describe semantic artefacts;
\item \textit{standard vocabulary} - a well-known, widely shared or standard metadata vocabulary is used;
\item \textit{primary metadata} - the original semantic artefact metadata are preserved in the catalogue;
\item \textit{version metadata} - metadata for each distribution/version of the semantic artefact is available;
\item \textit{human readable} - metadata is visible in the user interface in a harmonised manner;
\item \textit{machine readable} - metadata is accessible via API or machine-supported formats.
\end{enumerate}

\subsubsection*{Openness (Op)}
The concept of being open from different perspectives. On the one hand, it concerns technical openness, referring to the metadata handled in the catalogue, the software used to run the catalogue, and the services and protocols used to access the metadata. This aligns with the EOSC perspective towards community-driven Open Science infrastructures, which adheres to the UNESCO recommendation for Open Science \cite{unesco_unesco_2021}. On the other hand, openness also refers to the social attitude of enabling anyone interested in depositing and helping govern the catalogue. The four features identified in this dimension are:

\begin{enumerate}[label=\alph*)]
\item \textit{fully open source system (OSS)} - based on open source software;
\item \textit{customised open source system (OSS)} - the catalogue is based on an open source software, but the customised instance is not available for the public;
\item \textit{open model} - the metadata model/ontology used to document the semantic artefacts is openly available;
\item \textit{open contribution} - external or registered users can add/propose new semantic artefacts for inclusion.
\end{enumerate}

\subsubsection*{Quality (Qu)}
The possibility of having mechanisms to check the quality of the metadata provided and, thus, the catalogue itself. In particular, if processes and workflow are in place for peer reviewing new entities and curating the catalogue. The five features identified in this dimension are:

\begin{enumerate}[label=\alph*)]
\item  \textit{curation by owner only} - changes (or new submissions) to the semantic artefact can only be conducted by the catalogue owner;
\item  \textit{curation by maintainer} - changes (or new submissions) to the semantic artefact can be made by the maintainers/curators of the artefact;
\item  \textit{certified maintainer} - the maintainers/curators of the semantic artefacts are certified and assigned by the catalogue owner;
\item  \textit{metadata by editor} - metadata is curated by a group of editors;
\item  \textit{metadata by system} - metadata is generated/curated by an assessment system.
\end{enumerate}

\subsubsection*{Availability (Av)}
It refers to the availability of the metadata and whether there are methods in place for guaranteeing privacy and access only to certain data due to legal or other contextual issues. The three features identified in this dimension are:

\begin{enumerate}[label=\alph*)]
\item \textit{no restriction} - no authentication methods provided; contents are freely available without restrictions;
\item \textit{multilinguality} - items are translated and available in several languages and/or the content of artefact is accessible in multiple languages;
\item \textit{moderated services} - some functionalities for accessing and modifying semantic artefacts are available only to registered users and content creators.
\end{enumerate}

\subsubsection*{Statistics (St)}
The availability of statistics referred to the catalogue (number of semantic artefacts handled, number of users, etc.) in time to measure the usage of the catalogue and its growth. The three features identified in this dimension are:

\begin{enumerate}[label=\alph*)]
\item \textit{catalogue statistics} - basic metrics about the metadata catalogue;
\item \textit{resource statistics} - metrics on each semantic artefact;
\item \textit{social metrics} - social metrics for semantic artefacts, e.g., stars, likes, and the number of contributors.
\end{enumerate}

\subsubsection*{PID (Pi)}
The use of persistent identifiers (PIDs) that refer to the metadata of the various semantic artefacts described in the catalogue and the semantic artefacts themselves. The two features identified in this dimension are:

\begin{enumerate}[label=\alph*)]
\item \textit{PID in metadata} - PID used to identify the metadata values of the semantic artefact object included in the catalogue, e.g. ORCID for authors and ROR for research organizations;
\item \textit{resource PID} - PID used to identify the semantic artefact itself, e.g., DOI, PURL, w3id.
\end{enumerate}

\subsubsection*{Governance (Go)}
The rules that define the governance of the catalogue and its goals and purpose which should allow community input and responsibility for the integrity of the metadata. The three features identified in this dimension are:

\begin{enumerate}[label=\alph*)]
\item \textit{3rd party} - governance is supported by a 3rd party tool or platform, e.g., GitHub;
\item \textit{description} - governance is described as part of the catalogue;
\item \textit{rules} - rules for proposing new semantic artefacts to the catalogue are specified.
\end{enumerate}

\subsubsection*{Community (Co)}
The mechanism that is in place to involve the community in the catalogue, identifying and reaching target users' expectations and attracting stakeholders from diverse lived experiences and viewpoints. The four features identified in this dimension are:

\begin{enumerate}[label=\alph*)]
\item \textit{read only} - no direct involvement is possible; users can only communicate to the catalogue via read-only user interface, API or other mechanism such as email or feedback form;
\item \textit{read and write} - curators and developers can use services to get information from the catalogue and be directly involved through the creation of their records to be added to the catalogue and increase their visibility;
\item \textit{3rd party} - community features delegated to 3rd party tool or platform, e.g. GitHub issues;
\item \textit{suggestion} - a dedicated page for content suggestion (e.g. term change and term proposal) is available.
\end{enumerate}

\subsubsection*{Sustainability (Su)}
The models to sustain services financially and preserve the catalogue in the long run that are in place. The four features identified in this dimension are:

\begin{enumerate}[label=\alph*)]
\item \textit{organization} - the catalogue is a service provided by an organization (university, institute or one of its research units);
\item \textit{community} - a community with members from various organizations or infrastructure maintains the catalogue;
\item \textit{management board} - a multidisciplinary community-driven service strongly sustained by an operational team;
\item \textit{(research) project} - sustained by funds coming from one or more time-limited projects.
\end{enumerate}

\subsubsection*{Technology (Te)}
The tools that the catalogue should provide to enable users to have a better experience in exploring the data, such as REST APIs, Web search interfaces, SPARQL endpoints, etc. The four features identified in this dimension are:

\begin{enumerate}[label=\alph*)]
\item \textit{REST API} - a service to access semantic artefact information and/or metadata via a REST web service application programming interface;
\item \textit{web search GUI} - a service to access semantic artefact information and/or metadata via a web search or a graphical user interface;
\item \textit{SPARQL endpoint} - a service to access semantic artefact information and/or metadata via a SPARQL endpoint;
\item \textit{alignment} - a service that might be used within a catalogue to align (mapping) semantic artefacts and/or some of their parts (e.g. concepts, attributes, properties) one another or to deal with alignments/mappings without generating them.
\end{enumerate}

The identified features represent the aspects that can be directly assessed. Several other technical aspects on the backend solutions are generally not disclosed by the catalogue owners.

\subsubsection*{Transparency (Tr)}
The processes behind the governance of the catalogue, from the elections of new members of the various governing boards, curators, etc., to the clarity in exposing fees for the services offered by the catalogue and its revenue model. The three features identified in this dimension are:

\begin{enumerate}[label=\alph*)]
\item \textit{documented curation} - data flow of curation is documented;
\item \textit{automatic curation} - curation process happened automatically based on a documented process flow;
\item \textit{resource versioning} - records on previous versions of items are available.
\end{enumerate}

\subsubsection*{Assessment (As)}
The presence of some practice in place for assessing the catalogue against all these dimensions or additional ones, e.g. by adopting self-assessment exercises and/or by asking third parties to run an independent assessment of the catalogue. The two features identified in this dimension are:

\begin{enumerate}[label=\alph*)]
\item \textit{shared metrics} - assessment in terms of FAIRness is provided;
\item \textit{custom metrics} - assessment against catalogue's own assessment metrics.
\end{enumerate}

\subsection*{Catalogues assessment}
\label{sec:result-assessment}
The selected 26 catalogues of semantic artefacts are reported as follows:
\begin{itemize}
\item RDA Registry\cite{phipps2017-rdaregistry}, \url{https://www.rdaregistry.info/};
\item Architettura della Conoscenza (ARCO)\cite{carriero2019arco}, \url{http://wit.istc.cnr.it/arco};
\item BioPortal\cite{noy2009bioportal}, \url{https://bioportal.bioontology.org/};
\item TIB Terminology Service (TS4TIB), \url{https://service.tib.eu/ts4tib/};
\item Archivo\cite{frey2020dbpedia-archivo}, \url{https://archivo.dbpedia.org/list};
\item Linked Open Vocabularies (LOV)\cite{vandenbussche2017lov}, \url{https://lov.linkeddata.es/};
\item Prefix.cc, \url{https://prefix.cc/};
\item EU-Vocabularies, \url{https://op.europa.eu/en/web/eu-vocabularies/};
\item Ontology Design Patterns (ODP), \url{http://ontologydesignpatterns.org/};
\item Semantic Publishing and Referencing Ontologies (SPAR)\cite{peroni2018spar}, \url{http://www.sparontologies.net/};
\item FAIR Sharing\cite{sansone2019fairsharing}, \url{https://fairsharing.org/};
\item AgroPortal\cite{jonquet_2018}, \url{http://agroportal.lirmm.fr/};
\item Joint Food Ontology WG (JFOW), \url{https://github.com/FoodOntology/joint-food-ontology-wg};
\item Open Biological and Biomedical Ontology Foundry (OBO Foundry)\cite{smith2007obo}, \url{https://obofoundry.org/};
\item Bartoc, \url{https://bartoc.org/};
\item EMBL-EBI Ontology Lookup Service (EBI OLS) \cite{jupp2015new}, \url{https://www.ebi.ac.uk/ols/index};
\item International Virtual Observatory Alliance (IVOA), \url{https://ivoa.net/rdf/};
\item MatPortal, \url{https://matportal.org/};
\item EcoPortal\cite{kechagioglou2021ecoportal}, \url{https://ecoportal.lifewatch.eu/};
\item Loterre, \url{https://www.loterre.fr/};
\item MedPortal, \url{http://medportal.bmicc.cn/};
\item ESIP Community Ontology Repository, \url{http://cor.esipfed.org/};
\item NERC Vocabulary Service (NERC NVS), \url{http://vocab.nerc.ac.uk/};
\item Ontobee\cite{xiang2011ontobee}, \url{https://ontobee.org/};
\item HeTOP\cite{grosjean2011health}, \url{https://www.hetop.eu/hetop/};
\item Ontohub\cite{codescu2017ontohub}, \url{https://github.com/ontohub}.
\end{itemize}

Hereby, we provide the result of our assessment of the catalogues against the dimensions and related features described in the previous section in Table \ref{tab:results}. In addition to such an assessment, we also tracked the catalogue type. Specifically, we look into catalogues containing data and metadata, which we classified as a \textit{repository} according to the classification from Jonquet \cite{jonquet_2019}. These catalogues are marked with an asterisk (*) in the table header alongside the catalogue names. 

In addition, it is worth mentioning that the catalogues listed in Table \ref{tab:results} are not the same type of system. Indeed, some are ontology lookup services/repositories (e.g. AgroPortal and the EBI Ontology Lookup Service), while others are wider registries of standards/databases/policies (e.g., FAIRsharing) that include, but are not limited to, ontologies or other semantic artefacts. As such, their original intended scope and functionalities are inevitably different. However, according to the definition provided in this work, they all fall within the concept of catalogues of semantic artefacts.

The raw data of Table \ref{tab:results} are available online\cite{busse_raw_2023}.

\FloatBarrier
\begin{table}[t]
\caption{An overview of the dimensions and related features identified, accompanied by an assessment of 26 selected catalogues of semantic artefacts against such dimensions and features. The asterisk in the name of the catalogue means that it also stores a copy of the semantic artefacts it describes.}
\centering
\resizebox{\textwidth}{!}{%
\begin{tabular}{cl cccccccccccccccccccccccccc}
\toprule
    \multicolumn{2}{c}{\textbf{Dimension \& Features}} & 
    
    \rot{RDA Registry} & 
    \rot{ARCO*} & 
    \rot{BioPortal*} & 
    \rot{TS4TIB*} &
    \rot{Archivo*} &
    
    \rot{LOV*} & 
    \rot{Prefix.cc} & 
    \rot{EU-Vocabularies*} & 
    \rot{ODP*} &
    \rot{SPAR*} &
    
    \rot{FAIR Sharing} & 
    \rot{AgroPortal*} & 
    \rot{JFOW*} & 
    \rot{OBO Foundry} &
    \rot{Bartoc*} &
    
    \rot{EBI-OLS} & 
    \rot{IVOA*} & 
    \rot{MatPortal*} & 
    \rot{EcoPortal* } &
    \rot{Loterre*} &
    
    \rot{MedPortal*} & 
    \rot{ESIP*} & 
    \rot{NERC NVS} &
    \rot{Ontobee} &
    \rot{HeTOP} &
    \rot{Ontohub*}
    \\ 
\midrule


 

\multirow{6}{2em}{Me} 

 & custom vocabulary    & &x&x&x&x &x& & x& &x &x&x& &x&  &x&x&x&x&x &x&x&x&x&x &x \\
 & standard vocabulary  & & & & &  & & &x& &  & &x& & &  &x& & &x&x & & & & &  & \\
 & primary metadata     &x& &x&x&  &x& &x& &  &x&x& & &  & & &x&x&  &x& & & &  & \\
 & version metadata     & & &x& &x & & & & &  & &x& & &  & & &x&x&  &x& & & &  & \\
 & human readable       & & &x& &  & & & &x&x &x&x& &x&x & &x&x&x&  &x&x&x&x&  &x \\
 & machine readable     & & &x& &  &x& &x& &  &x&x& &x&x & & &x&x&  &x&x&x& &  & \\
 
\midrule

\multirow{4}{2em}{Op} 

 & fully OSS         & & &x& &x &x&x& &x&x & &x& &x&x & & &x&x&  & & &x&x&  &x \\
 & customised OSS    & & & &x&  & & & & &  & & & & &  &x& & & &  & &x& & &  & \\
 & open model        & &x& & &x & & & & &  &x&x& & &  & & & &x&x & & &x& &x &x \\
 & open contribution & & &x& &x &x&x&x&x&x &x&x& &x&  & &x&x&x&x & &x&x& &  &x \\
 
\midrule

\multirow{5}{2em}{Qu} 

 & curation by owner only & &x& &x&  & & & & &  & & & & &  &x& & & &  &x& & & &  & \\
 & curation by maintainer & & &x& &  &x& &x&x&x &x&x&x&x&  &&x&x&x&x  & & & & &  & \\
 & certified maintainer   &x& & & &  & & & & &  &x& & & &  & & & & &  & & & & &  & \\
 & metadata by editor     & & & & &  & & & & &  &x&x& & &x & & & &x&  & & & & &  & \\
 & metadata by system     & & & & &x & & & & &  & &x& & &  & & & &x&  & & & & &  & \\
 
\midrule

\multirow{3}{2em}{Av} 

 & no restrictions      &x&x&x&x&x &x&x&x&x&x &x&x&x&x&x &x&x&x&x&x &x&x&x&x&  &x \\
 & multilinguality      &x& & & &  & & &x& &  & &x& & &  & & & & &x & & & & &x &  \\
 & moderated services   & & &x& &  & & & & &  &x&x& & &  & & &x&x&  &x& &x& &x &  \\
 
\midrule

\multirow{3}{2em}{St} 

 & catalogue statistics   &x& &x&x&x &x& & & &  &x&x& & &  & & &x&x&  &x& & & &  &x \\
 & resource statistics  &x& &x&x&x &x& &x& &  & &x& & &  & & &x&x&x &x& & &x&x &  \\
 & social metrics       & & &x& &  & & & & &  & &x& &x&  & & &x&x&  & &x& & &  &  \\
 
\midrule

\multirow{2}{2em}{Pi} 

 & PID in metadata   & &x& & &  &x& & & &  &x&x& &x&  &x& & & &  & & x&x & &  &  \\
& resource PID           & & & & &  & & & & &x & & & &x&  & & & & & x& & & & &  &  \\

\midrule

\multirow{4}{2em}{Go} 
 & 3rd party &x&&&& &&&&& &&&x&x& &&&&& &&&&& & \\
 & description &&x&x&&x &x&&x&x& &x&x&&x& &x&x&x&x&x &&&&x& & \\
 & rules &&&x&&x &x&&x&x&x &x&x&&x& &&&x&x&x &&x&&& & \\
 
\midrule

\multirow{4}{2em}{Co} 
 & read only        &&x&&x& &x&&x&& &&&&&x &x&&&&x &&&&&x & \\
 & read and write   &&&x&& &&x&&x& &x&x&&& &&&x&x& &x&&&& & \\
 & 3rd party        &x&&&& &&&&x&x &&&x&x& &x&&&& &&&&& &x \\
 & suggestion       &&&x&&x &&&&& &x&x&&& &&x&x&x& &x&x&x&x& & \\
 
\midrule

\multirow{4}{2em}{Su} 

 & organization         &x&x&x&x&x & &x&x&x&  &x&x& & &x &x& &x&x&x &x& &x&x&x &  \\
 & community            & & & & &  &x& & & &x & & & &x&x & & & & &  & & & & &  &x \\
 & management board     & & & & &  & & & &x&  &x& & &x&x & & & & &  & &x& & &  &x \\
 & (research) projects  & & &x& &  & & & & &  &x&x& & &  & & &x&x&  & &x& & &  &  \\
 
\midrule

\multirow{3}{2em}{Te} 
 & REST API             & & &x&x&x &x&x&x& &  &x&x& & &x &x& &x&x&x &x&x&x& &x &  \\
 & web search GUI       &x&x&x&x&x &x&x&x&x&  &x&x& &x&x &x&x&x&x&x &x&x&x&x&x &x \\
 & SPARQL endpoint      & &x& & &  & & &x& &  & &x& & &  & & &x&x&x &x&x&x&x&  & \\
 & alignment            & & &x& &  & & &x& &  & &x& & &  & & &x&x&  &x& & & &  & \\
 
\midrule

\multirow{3}{2em}{Tr} 

 & documented curation  &x& & & &x & & &x&x&x &x&x& &x&  & &x& & &  & & & & &  &x \\
 & automatic curation   & & & & &x & & & & &  & &x& & &  & & & & &  & & & & &  &  \\
 & resource versioning  & & &x& &x & & &x& &  &x&x& &x&  & & &x&x&  &x&x& & &  &x \\
 
\midrule

\multirow{2}{2em}{As} 

 & shared metrics   & & & & & x & & & & &  & &x& & &  & & & &x &  & & & & &  &  \\
 & custom metrics   & & & & &  & & & & &  & & & &x&  & & & & &  & & & & &  &  \\
 


 
\bottomrule

\end{tabular}}
\label{tab:results}
\end{table}
\FloatBarrier

\section*{Discussion}
\label{sec:discussion}

This section contains an analysis of the maturity dimensions based on the assessment results outlined in Table \ref{tab:results}. Maturity dimensions are examined in the order they appear in the table to summarize the current status of implementation regarding each category.

\textit{Metadata (Me).} Adopting standard vocabulary for metadata is an essential aspect of joining, comparing, and curating semantic artefacts. In addition, it fosters the interoperability of these items. Yet, only 19\% (5) use standard vocabularies. Further analysis is needed to confirm that all the other catalogues do not use standard vocabularies. Indeed, these catalogues may use custom vocabularies made by extending standard vocabularies.

\textit{Openness (Op) and Quality (Qu).} All the catalogues maintained by a community with members from different organizations/infrastructures (5 out of 26) are based on open-source tools and provide no authentication methods/restrictions to their contents. Furthermore, catalogues that enable external/registered users to add/propose new semantic artefacts to include in the catalogue (65\%) delegate the quality control of these changes to the maintainers/curators of the artefacts. Indeed, generally, more than half of the analysed catalogues (65\%) permit the curation of their data by either the owners of the catalogue or the maintainers of the semantic artefacts. No catalogue provides both strategies. This aspect is a positive sign, which shows that most of the catalogues are concerned with guaranteeing a good quality of their data.

\textit{Availability (Av) and PID (Pi).} All catalogues, with only one exception, provide access with no restriction. However, 8 (i.e., around 31\%) include functionalities, mostly APIs, accessible only to registered users. Only a tiny percentage of catalogues (i.e., 11.5\%) use PIDs for identifying resources. At the same time, a more significant number of catalogues (i.e., 30.8\%) provide a PID for the metadata of a given record. They are typically used to represent the creator, the contributor or for more specific aspects, e.g., the URI of imported ontologies. In only one case, the catalogue provides PIDs for both the identification of resources and their metadata; most of the time, the catalogues do not assume the role of assigning and managing PIDs to their resources.

\textit{Statistics (St).} Half of the catalogues provide resource statistics, typically including the number of classes, properties and axioms included in their semantic artefacts. Most of these last (corresponding to around 35\% of the total) also provide general statistics that aggregate information across all resources. In a few cases, social statistics are enabled by the catalogue or included from another source, e.g., in the form of metrics taken from GitHub metadata where the original resources are stored (e.g., number of received stars and contributors).

\textit{Governance (Go) and Community (Co).} Governance processes and/or structure are described by over half of the catalogues, 73\% of which explicitly provide rules for contributors willing to propose new resources. In 7 cases, external 3rd-party solutions, particularly GitHub, are used as resource management tools. Some of the catalogues are more open to the community's contribution than others. In particular, while almost one-third only provide the capabilities to communicate with the catalogue through read-only APIs, only 8 out of 26 also provide the possibility for resource creation.

\textit{Sustainability (Su).} In 2 cases, the sustainability could not be assessed. In the remaining catalogues, the sustainability seems sufficiently stable since they are maintained by at least one organization, community, or management board. Furthermore, six of them are also supported by research projects.

\textit{Technology (Te).} Most catalogues (92\%) provide at least a web search GUI. The non-SPARQL catalogues represent 61\% (16) of the total; around 88\% of these use a web search GUI, and 56\% combine it with REST APIs. Therefore, only 10 out of 26 provide a SPARQL endpoint. 70\% of these provide machine-readable metadata. Overall, the preferred technology used by all the catalogues integrating machine-readable metadata is either REST APIs, web search GUI, or both. Many catalogues incorporate all three technologies (30\%), i.e. REST API, web search GUI, and SPARQL endpoint, especially the catalogues based on the OntoPortal software \cite{jonquet_2023}. These catalogues provide data with no authentications/restrictions, and contents are freely available without restrictions. In addition, for most of them, external/registered users can add/propose new semantic artefacts to be included.

\textit{Transparency (Tr).} Only 10 catalogues (i.e., around 38\%) explicitly document a workflow for data curation, which in two cases is mainly automated; 9 of these are open for modifications to be made to the semantic artefacts by external/registered users. In 11 cases, the catalogue also provides previous resource versions, enabling a versioning system useful for backward compatibility and documentation.

\textit{Assessment (As).} Most catalogues do not provide information on (self-)assessment against quality criteria. Among the few exceptions, AgroPortal \cite{jonquet_2018} includes an assessment in terms of FAIR score, an evaluation of the satisfaction of each aspect of the FAIR principles. This FAIRness assessment methodology \cite{amdouni_23} has been recently transposed to 6 other OntoPortal-based catalogues, including solely EcoPortal, in our study. FAIR score includes many questions specific to ontologies and semantic resources, with the ability to compute the score for each resource and the whole catalogue. On the other hand, Archivo \cite{frey2020dbpedia-archivo} proposes a rating based on a set of automatically assessed criteria (whether the ontology is retrievable and parsed correctly, is provided with a clear and proper license statement, and is logically consistent).  

The analysis showed the current state of the identified catalogues, and it can also serve as a guidance for shaping future developments. The maturity model created was a first attempt toward understanding catalogues of semantic artefacts. In addition, the catalogue assessment we performed was a preliminary (and we believe a successful) attempt to perceive the effectiveness of the maturity model. 

The analysis will be complemented in the future with the ongoing work on minimum metadata sets and interoperability indicators currently carried out within the EOSC Task Force on Semantic Interoperability. By combining these efforts, the aim is to provide recommendations for governance and processes for preserving and maintaining semantic artefacts. This will involve identifying gaps and improvement areas in semantic interoperability approaches and developing strategies to address these challenges.

In the future, we aim to interlink aspects of the dimensions identified within the maturity model presented in this paper with recommendations of other EOSC Task Forces. For instance, the ESOC Task Force on FAIR Metrics and Data Quality has produced some guidelines\cite{lacagnina_towards_2023,wilkinson_community-driven_2022,wilkinson_fair_2022} that may be used in the context of the \emph{Quality} maturity dimension for providing even more in-depth specifications for measuring the FAIRness of semantic artefacts catalogues. Similarly, the work under development in the EOSC Task Force on PID Policy and Implementation (\url{https://www.eosc.eu/advisory-groups/pid-policy-implementation}) may provide additional insights related to the \emph{PID} maturity dimension highlighted in this article. 

Another aspect that would deserve further study concerns the relation of our dimensions with those described in other maturity models available in the literature, such as some of those we used for defining the dimensions. Indeed, it would be essential to make maturity models semantically interoperable since it may also facilitate further analysis of the systems they measure. We will leave this investigation for future developments.

The maturity dimensions, the related features, and the list of assessed catalogues are open for extensions and refinements in the future, especially in light of the feedback from the community. For instance, a few people suggested, via their reviews (\url{https://prereview.org/preprints/doi-10.48550-arxiv.2305.06746}), the addition of further perspectives to include in the maturity model, such as metrics measuring the facility to find specific contextual information in a catalogue, the size of the community using it, its Technological Readiness Level (TRL), its time of existence, and the specification of the types of semantic artefacts handled by a catalogue with appropriate dimensional features associated to each type. Others have already suggested the inclusion of additional catalogues in the assessment. While these are all valid suggestions, further technical analyses are necessary for a future refinement/extension of the maturity model. While currently, within the Task Force, there are no particular plans for adding more technical/operational features, being the data and resources produced in this work available on Zenodo \cite{corcho_catalogues_2023,busse_raw_2023} will allow and enable the community to create further versions of them if needed, even outside the framework of the EOSC Task Forces.

We believe that having catalogues of semantic artefacts compliant with the dimensions and features we identified in the maturity model is a crucial step towards the broad adoption and reuse in research, by all the involved users (researchers, data stewards, etc.), of the semantic artefacts they provide. Indeed, without sound and well-governed catalogues, we do not have persistence, availability and maintenance of semantic artefacts in the long term; and, without guaranteeing these activities, it is challenging to build users’ trust in such artefacts, which is crucial to enable and push their broad adoption and reuse in the EOSC and, more generally, in research. However, we recognise a perceived need by the community to extend, in the future, the effort of the EOSC Task Force on Semantic Interoperability to focus on: (a) understanding how semantic artefacts should be used productively in and for research; (b) how data stewards can efficiently and effectively use these artefacts in metadata specification and metadata curation workflows; and (c) how to bring data producers, publishers, researchers, to adopt them.

\section*{Methods}
\label{sec:methods-material}

This section presents the methodology followed to identify the dimensions and related features of the maturity model and analyse the catalogues of semantic artefacts. The process was divided into five steps, described in the following subsections.

\subsection*{Analysis of existing literature}
We involved all the members of the EOSC Task Force on Semantic Interoperability, which included several experts on the topic with very heterogeneous backgrounds, to provide us with relevant material related to at least one of the two aspects of interest for our study, i.e., (1) definitions of catalogues of semantic artefacts and (2) dimensions that can be used to measure the maturity of such catalogues. 

After gathering all such relevant documents, we asked the members of the Task Force to read them and highlight any passage in the text referring either to definitions related to catalogues of semantic artefacts or maturity measures. We also asked that the reading and analysis of each document must be performed, when possible, by a member who did not propose it to avoid possible biases. 

We found that 15 of the gathered documents contained relevant texts, as shown in Table \ref{t_mat_dimensions}. The raw data of this overview are available online\cite{busse_raw_2023}. The authors analysed all the quotations highlighted to gather maturity dimensions. In particular, the names of the dimensions were compiled from the analysed literature, thus reusing a terminology already shared with the scientific community and heterogeneously distributed in different documents. The 12 dimensions extracted, which we used as a starting point for identifying additional features characterising them, are summarised in Table \ref{t_mat_dimensions}.

\subsection*{Collection of catalogues}
A preliminary search was conducted to collect potential catalogues of semantic artefacts. Potential catalogues of semantic artefacts have been identified by direct knowledge of the co-authors and the other members of the Task Force.

The authors then screened the resulting list of potential catalogues to remove duplicates and those irrelevant to the study. In particular, we decided to keep only those potential catalogues that refer mainly to semantic artefacts in the analysis. This exclusion criterion has been made to filter out (i) generic repositories that may also contain semantic artefacts, even if it is not the primary resource types they refer to (e.g., Zenodo), and (ii) generic repositories (e.g., Google or other general-purpose search engines). The resulting set included 26 selected catalogues. With this list, our goal was not to be exhaustive but to cover multiple application domains well and get a good representation of the underlying technology used to build the catalogues (e.g., OntoPortal, OLS, SKOSMOS, etc.).

\subsection*{Setup of catalogues assessment}
The identified catalogues were evaluated based on their relevance to semantic artefacts. A spreadsheet was created with the selected catalogues listed on rows and the 12 maturity dimensions on columns. Additional columns were dedicated to the authors' names (acting as reviewers) and comments. Each reviewer was assigned some catalogues to review.

Twelve separate tabs in the spreadsheet described the possible features characterising each maturity dimension. These tabs contained the name of the dimension and a set of particular features, each with a number, a short name, a description, the name of the reviewer that proposed it, and whether the group had validated it. In addition, there was a column for possible comments.

The main table in the first tab was extended with two additional columns. The first column allowed us to specify whether the catalogue under analysis stores a copy of the semantic artefacts it describes. The second column referred to the software or tool used to implement the catalogue to distinguish if the software adopted was generic (i.e., can be used to deploy multiple catalogues) and open-source.

\subsection*{Analysis}
Each reviewer evaluated a first small set of 2-3 catalogues against the 12 dimensions identified before, which we considered fixed. For each catalogue and each dimension, the reviewer had to select which features of the dimensions applied to the catalogue. If, for a given dimension, a particular feature was not already present, the reviewer could add such a new feature, adding a name and a description for it and making it available for the following reviewers. Several features for each dimension (from 3 to 7), have been added during this step, for a total of 63 potential features.

After the first analysis, early results and issues were discussed by all reviewers. The analysis was then extended by assessing other potential catalogues, i.e., an additional five for each reviewer, to have a clear view of other aspects that did not arise from the former analysis. In addition, we decided to invite other experts (external to the Task Force) to such a study afterwards, ideally after we have assessed all the potential catalogues. Our perceived risk here was to be biased by a particular point of view while we were still in a phase of preliminary analysis.

\subsection*{Harmonization and summarization}\label{sec:harmonization}
After completing the catalogue descriptions, a final review and summarization were conducted. Each reviewer was tasked with analyzing 2 to 4 dimensions to review the corresponding features provided by other reviewers and suggested potential edits such as merging similar features, removing irrelevant ones, or splitting them into separate aspects. As a result, 20 out of 63 identified features have been removed from the analysis or merged with others, keeping 43 final features. The main table was updated accordingly to reflect any changes made by the reviewer.

A final meeting was held to review and harmonize the catalogue dimensions and features to ensure a consistent set of features that could be used to compare and analyze the different catalogues. During this meeting, any remaining inconsistencies or ambiguities in the features were discussed, and final decisions were made on harmonising them across all catalogues.

\section*{Data Availability}
All data gathered and created in this work have been uploaded on Zenodo. In particular, the maturity model (i.e., the dimensions and the related features) described in this article is available in three different formats (PDF, CSV, XSLX) at \url{https://doi.org/10.5281/zenodo.10625936}\cite{corcho_catalogues_2023}. In addition, the raw data with the quotations from the 15 selected documents referring to maturity dimensions and the machine-readable version of the data in Table \ref{tab:results} are available in two different formats (CSV, XSLX) at \url{https://doi.org/10.5281/zenodo.10618181}\cite{busse_raw_2023}.

\section*{Code Availability}
Not applicable.

\bibliography{paper}

\section*{Acknowledgements}

The authors thank all members of the EOSC Task Force on Semantic Interoperability and all the reviewers of the prior version of this article (five anonymous experts involved in the single-blind reviewing process of the 22nd International Semantic Web Conference and Scientific Data, plus Allyson L. Lister, Andrea Scharnhorst, Hagar Lowenthal, and Lars Vogt, who provided open reviews at PREreview.org - \url{https://prereview.org/preprints/doi-10.48550-arxiv.2305.06746}) for the fruitful discussions, suggestions, and joint work. The work of SP has been partially funded by the European Union's Horizon 2020 research and innovation program under grant agreement No 101017452 (OpenAIRE-Nexus) and the European Union’s Horizon Europe research and innovation program under grant agreement No 101095129 (GraspOS). The work of CJ has been partially funded by the European Union's Horizon Europe research and innovation program under grant agreement No 101057344 (FAIR-IMPACT). The work of AM has been partially funded by the Data Repository Platform project (ARP) of the Hungarian Research Network (HUN-REN), with SZTAKI as its Leading Partner, taking place in SZTAKI DSD.

\section*{Author contributions statement}
Authors’ contribution according to CRediT (\url{https://credit.niso.org/}): Data Curation, Investigation, Methodology, Visualization (all authors); Conceptualization, Supervision (OC, SP); Validation (FE, AM, SP, ES); Writing – original draft, Writing – review \& editing (FE, IH, CJ, AM, SP, ES).

\section*{Competing interests}
All the authors are members of the EOSC Task Force on Semantic Interoperability, except CJ and IH.

\end{document}